\documentclass[twocolumn]{aastex62}
\usepackage{graphics,epsf}
\usepackage{amsmath}                % American Mathematical Society package
\usepackage{amsfonts}               % American Mathematical Society fonts
\usepackage{amssymb}                % American Mathematical Society symbol
\usepackage{epsfig}                 % EPS figures
\usepackage{graphicx}
\usepackage{float}
\usepackage{color}
\usepackage[para,online,flushleft]{threeparttable}

\newcommand{\cm}{{~\rm cm}}
\newcommand{\km}{{~\rm km}}
\newcommand{\s}{{~\rm s}}

\newcommand{\yr}{{~\rm yr}}

\newcommand{\Gyr}{{~\rm Gyr}}
\newcommand{\pc}{{~\rm pc}}

\newcommand{\days}{{~\rm days}}

\begin{document}

\title{Common envelope to explosion delay time of type Ia supernovae}

%% \correspondingauthor{Noam Soker}
%% \email{soker@physics.technion.ac.il}

%% \author{Efrat Sabach}
%%% \affiliation{Department of Physics, Technion, Haifa, 3200003, Israel}

\author[0000-0003-0375-8987]{Noam Soker}
%\affil{Departmeמt of Physics, Technion, Haifa 3200003, Israel}
\affiliation{Department of Physics, Technion, Haifa, 3200003, Israel; soker@physics.technion.ac.il}
\affiliation{Guangdong Technion Israel Institute of Technology, Shantou 515069, Guangdong Province, China}

%%% \author{Ran Segev\altaffilmark{1}, Efrat Sabach\altaffilmark{1}, Noam Soker\altaffilmark{1,2}}

%%% \altaffiltext{1}{Department of Physics, Technion -- Israel Institute of Technology, Haifa 32000, Israel; efrats@physics.technion.ac.il; soker@physics.technion.ac.il}
%%% \altaffiltext{2}{Guangdong Technion Israel Institute of Technology, Shantou, Guangdong Province 515069, China}
   
\begin{abstract}
I study the rate of type Ia supernovae (SNe Ia) within about a million years after the assumed common envelope evolution (CEE) that forms the progenitors of these SNe Ia, and find that the population of SNe Ia 
with short CEE to explosion delay (CEED) time is $\approx {\rm few} \times 0.1$ of all SNe Ia. I also claim for an expression for the rate of these SNe Ia that occur at short times after the CEE, $t_{\rm CEED} \la 10^6 \yr$, that is different from that of the delay time distribution (DTD) billions of years after star formation. This tentatively hints that the physical processes that determine the short CEED time  distribution (CEEDTD)  are different (at least to some extend) from those that determine the DTD at billions of years. 
To reach these conclusions I examine SNe Ia that interact with a circumstellar matter (CSM) within months after explosion, so called SNe Ia-CSM, and the rate of SNe Ia that on a time scale of tens to hundreds of years interact with a CSM that might have been a planetary nebula, so called SNe Ia inside a planetary nebula (SNIPs).
I assume that the CSM in these populations results from a CEE, and hence this study is relevant mainly to the core degenerate (CD) scenario, to the double degenerate (DD) scenario, to the double detonation (DDet) scenario with white dwarf companions, and to the CEE-wind channel of the single degenerate (SD) scenario.  
\\
\textit{Keywords:} (stars:) white dwarfs -- (stars:) supernovae: general -- (stars:) binaries: close
\end{abstract}

%Keywords: (stars:) white dwarfs, (stars:) supernovae: general

% ==========================================================
\section{INTRODUCTION}
\label{sec:intro}
% ==========================================================

Type Ia supernova (SN Ia) research has advanced in recent years due to new observations and theoretical models. 
%%% , as well as sociology. 
At present there is no consensus on the scenarios that bring white dwarfs (WDs) to experience thermonuclear explosions as SNe Ia (for recent reviews see  \citealt{Maozetal2014, LivioMazzali2018, Wang2018, RuizLapuente2019}, in particular \citealt{Soker2018Rev} for a table comparing the five scenarios). 
For that I list (in an alphabetical order) \textit{all} binary scenarios and emphasise the differences between them.  
%%% (rather than mentioning only the two scenarios that were popular in the literature in the previous millennium).  

(1) In the \textit{core-degenerate (CD) scenario} a CO WD companion merges with the CO (or possibly HeCO) core of a massive asymptotic giant branch (AGB) star during a common envelope evolution (CEE). The CD scenario is a separate scenario (e.g., \citealt{Kashi2011, Ilkov2013, AznarSiguanetal2015}) because (a) at explosion there is one star, (b) it leaves no remnant, and (c) the delay time from CEE to explosion is set by the evolution of a single WD remnant of the merger.  

(2) In the \textit{double degenerate (DD) scenario} two WDs merge (e.g., \citealt{Webbink1984, Iben1984}), most likely in a violent process (e.g., \citealt{Pakmoretal2011, Liuetal2016}) a long time after the CEE. 
One major open parameter in the DD scenario is the  time delay from merger to explosion (merger explosion delay, or MED; e.g., \citealt{LorenAguilar2009, vanKerkwijk2010, Pakmor2013, Levanonetal2015}).
This is a separate scenario because (a) at explosion or shortly before explosion there are two WDs, (b) the explosion leaves no remnant, and (c) the delay time from CEE to explosion is set mainly by gravitational wave radiation of the two WDs. 
Note that the two WDs need not be CO WDs, e.g., one of them might be a helium WD or a HeCO hybrid WD (e.g., \citealt{YungelsonKuranov2017, Zenatietal2019}).

(3) In the \textit{double-detonation (DDet) mechanism} the companion transfers mass to a CO WD, and the ignition of the helium-rich layer that was accreted from a companion ignites the CO WD (e.g., \citealt{Woosley1994, Livne1995, Shenetal2018}). This is a separate scenario because (a) there are two stars at explosion where only one of them explodes and (b) it leaves a surviving star, either an evolved helium star or a WD. Although there might be two WDs at explosion, because one WD survives the explosion this scenario is different than the DD scenario. 
In many channels of the DDet scenario the system experiences a CEE to bring closer the helium-rich companion and the CO WD. 

(4) In the \textit{single degenerate (SD) scenario} a WD accretes a hydrogen-rich material from a non-degenerate companion. The WD reaches close to the Chandrasekhar mass limit ($M_{\rm Ch}$), and explodes (e.g., \citealt{Whelan1973, Han2004, Wangetal2009}), either as soon as it reaches this mass or much later after it loses some of its angular momentum (e.g., \citealt{Piersantietal2003, DiStefanoetal2011, Justham2011}).
This scenario is different than the other scenarios by that the WD reaches a mass of $\simeq M_{\rm Ch}$ by accreting hydrogen-rich gas. If the explosion takes place after a long delay, it might leave behind a subdwarf B star or a WD. In the CEE-wind SD scenario that \cite{MengPodsiadlowski2017} suggested, the explosion might occur shortly after a CEE if the WD is a hybrid CONe WD \citep{MengPodsiadlowski2018}. In this scenario, that \cite{MengPodsiadlowski2018} predict to be $\approx 5-10 \%$ of all SNe Ia,  the WD accretes hydrogen-rich material from the envelope of a giant star while it spirals-in and ejects the envelope. This channel of the SD scenario is relevant to the present study.    

(5) The \textit{WD-WD collision (WWC) scenario} involves the collision of two WDs at about their free fall velocity into each other (e.g., \citealt{Raskinetal2009, Rosswogetal2009, Kushniretal2013, AznarSiguanetal2014}). 
\cite{Toonenetal2018} conduct a thorough population synthesis study and conclude, as some earlier studies did, that the SN Ia rate from the WWC scenario might be of the order of $0.1 \%$ of all SNe Ia. Follow up studies reach the same qualitative conclusion (e.g., \citealt{HallakounMaoz2019, HamersThompson2019}).
As this scenario seems incapable to explain even a small fraction of SNe Ia, and it does not need the CEE, I will not consider it anymore in the present study. 

A different classification that is more relevant to the explosion mechanism and the nucleosynthesis yield is to WDs that explode with masses near the Chandrasekhar mass limit, `$M_{\rm Ch}$ explosions`, and WDs that explode with masses below that mass, `sub-$M_{\rm Ch}$ explosions` (e.g., \citealt{Maguireetal2018}). Crudely, the DD, DDet, and WWC scenarios belong to the sub-$M_{\rm Ch}$ explosions while the CD and the SD scenarios belong to $M_{\rm Ch}$ explosions.
As there are indications for $M_{\rm Ch}$ explosions  
(e.g.,  \citealt{Ashalletal2018, Dhawanetal2018, Diamondetal2018}) there is place to consider the CD and the SD scenarios. In earlier papers I already noted the severe difficulties with the SD scenario (e.g., \citealt{Soker2018Rev}). There are also strong indications from the behaviour of SNe Ia for the presence of sub-$M_{\rm Ch}$ explosions (e.g. \citealt{Scalzo2014b, Blondin2017, Goldstein2018, Wygoda2019a, LevanonSoker2019, KonyvesTothetal2019}). 
The DD scenario is a promising  sub-$M_{\rm Ch}$ scenario (e.g., \citealt{Soker2018Rev} for a review), possibly the hybrid DD scenario where there is one HeCO WD (e.g., \citealt{YungelsonKuranov2017, Zenatietal2019}).  

In many cases more than one scenario can account for a specific observation, and so it is mandatory to consider all relevant scenarios. For example, the presence of a circumstellar matter (CSM) is expected in some SNe Ia of all scenarios, beside the WWC scenario. A massive CSM with hydrogen is expected only in the CD scenario, in the DD scenario, and in the CEE-wind channel of the SD scenario. In the DD scenario this is the case if the two WD merge shortly after the CEE. For example, the CSM of SN Ia PTF11kx \citep{Dildayetal2012} is too massive for most channels of the SD scenario, and requires the CEE-wind channel of the SD scenario \citep{MengPodsiadlowski2017}, the CD scenario \citep{Sokeretal2013}, or with some fine tuning the DD scenario. 
%%% For that, new discoveries of SN Ia-CSM (e.g., \citealt{Grahametal2019}) should consider all relevant scenarios, and not only some channels of the SD scenario.  
     
In the present study I consider the common envelope to explosion delay (CEED) time of the CD and DD scenarios, some channels of the DDet scenario (those that experience the CEE), and the CEE-wind channel of the SD scenario. In section \ref{sec:delay} I define the CEED time in relation to the delay time distribution (DTD) from star formation to explosion, and the merger/accretion to explosion delay (MED) time. In section \ref{sec:Estimating CEED} I derive a crude expression to the CEED time  distribution (CEEDTD)  and discuss its implications, and in section \ref{sec:summary} I summarise this short study. 

%==========================================================
\section{The delay times}
\label{sec:delay}
% ==========================================================
%=========================
\subsection{Delay time distribution (DTD)}
\label{subsec:delay}
% ========================

The DTD is the distribution of the delay time  from star formation to the actual SN Ia explosion, $t_{\rm SF-E}$. Different studies with different techniques (see, e.g., \citealt{Heringeretal2019}) have deduced somewhat different expressions for the DTD from observations (e.g., \citealt{Grauretal2014, Heringeretal2017, MaozGraur2017}). The two recent studies of \cite{FriedmannMaoz2018} for the rate of SNe Ia in galaxy clusters and that of \cite{Heringeretal2019} for field galaxies derive very similar parameters in the expression for the  DTD   
\begin{equation}
\dot N_{\rm DTD} \equiv \left( \frac {d N_{\rm Ia}}{dt} \right)_{\rm DTD} = A \left( \frac{t}{1 \Gyr} \right)^{\alpha}. 
\label{eq:dotN}
\end{equation}
\cite{FriedmannMaoz2018} derive $A = 5-8 \times 10^{-13} M^{-1}_\odot \yr^{-1}$ and $\alpha=-1.30^{+0.23}_{-0.16}$, while \cite{Heringeretal2019}  derive $A = 7\pm 2 \times 10^{-13} M^{-1}_\odot \yr^{-1}$ and $\alpha=-1.34^{+0.19}_{-0.17}$.
I will use these results in what follows (but I note recent different results, e.g., \citealt{Frohmaieretal2019}).  
 
Some studies compare this derived DTD to the spiralling-in time due to gravitational wave emission of two WDs in the frame of the DD scenario, $t_{\rm GW}$. But it is important to remember that there are actually two other evolutionary phases that add up to yield the total delay time from star formation to explosion in the DD scenario, $t_{\rm SF-E}({\rm DD})$. These are the times from star formation to the formation of the two WDs in the post-CEE phase, $t_{\rm SF-CE}$, and the time from the merger of the two WDs to explosion, the MED time $t_{\rm MED}$ (section \ref{subsec:MED}).   Namely, 
\begin{equation}
t_{\rm SF-E}({\rm DD}) = t_{\rm SF-CE} + t_{\rm CEED} = t_{\rm SF-CE} + t_{\rm GW} + t_{\rm MED} , 
\label{eq:DDtSFE}
\end{equation}
where $t_{\rm CEED}$ is the time from the end of the CEE to explosion. 
  If both $t_{\rm SF-CE} \ll t_{\rm GW}$ and $t_{\rm MED} \ll t_{\rm GW}$ then the assumption $t_{\rm SF-E}({\rm DD}) \simeq  t_{\rm GW}$ holds. 
  
In the CD scenario the WDs merge during the CEE, and so   
 \begin{equation}
t_{\rm SF-E}({\rm CD}) = t_{\rm SF-CE} + t_{\rm CEED} = t_{\rm SF-CE} +  t_{\rm MED} . 
\label{eq:CDtSFE}
\end{equation}
I discuss the MED time in section \ref{subsec:MED} and the CEED time in section \ref{subsec:CEED time}.

\cite{FriedmannMaoz2018} fit their DTD down to delay time of $t_{\rm SF-E} = 1.5 \Gyr$ and consider SNe Ia to occur from 
$t_{\rm SF-E} = 0.04 \Gyr$ to present $t_{\rm SF-E} = 13.7 \Gyr$. 
They find a production efficiency (defined as Hubble-time-integrated SN Ia number per formed stellar mass) of $n_{\rm Ia}\simeq 0.003-0.008 M^{-1}_\odot$. 
\cite{Heringeretal2019} consider SNe Ia to occur in the time interval from  $t_{\rm SF-E} = 0.1 \Gyr$ to $t_{\rm SF-E} = 13.7 \Gyr$ and find $n_{\rm Ia} \simeq 0.003-0.006 M^{-1}_\odot$.   

As for the  slope of the DTD, \cite{Heringeretal2019} note that a slope of $\alpha \simeq -1.35$ falls between the expected value of the DD scenario and the DDet scenario (e.g., \citealt{Ruiteretal2011}). \cite{Neunteufeletal2019} argue that the DDet scenario with a non-degenerate helium donor can account for no more than few percent of all SNe Ia.  Indeed, \cite{Ruiteretal2011} find in their population synthesis study that most of their DDet SNe Ia come from WD donors. These systems experience a CEE phase, and are relevant to the present study. 
  
%=========================
\subsection{Merger to explosion delay (MED) time}
\label{subsec:MED}
% ========================

In an earlier study \citep{Soker2018Rev} I argued that in a large fraction of SNe Ia there must be a substantial time delay between the end of the merger of the WD with a companion or the end of mass accretion on to the WD and the terminal explosion of the WD as a SN Ia. 
Several observations suggest the existence of a merger/accretion to explosion delay (MED) time, $t_{\rm MED}$. I give here a brief summary before I introduce the motivation for my definituon of the CEED time (section \ref{subsec:CEED time}). 

(1) If the explosion of the two WDs in the DD scenario occurs as they dynamically interact, then the explosion is asymmetrical (e.g.,  \citealt{Kashyapetal2017, Pakmor2012, Tanikawaetal2015, vanRossumetal2016}), contradicting the structure of most SN Ia remnants (SNRs Ia) that tend to be spherical or axisymmetrical (e.g., \citealt{Lopezetal2011}). In that respect I note that a surviving WD companion in the DDet scenario also leads to a SNR Ia that possesses non-spherical morphological features (e.g., \citealt{Papishetal2015, Tanikawaetal2018, Tanikawaetal2019}).  

(2) Several SNe~Ia show early ($\la 5 \days$) excess emission in their light curve (e.g., \citealt{Marionetal2016, Hosseinzadehetal2017, Shappee2019, Dimitriadisetal2019a,  Jiang2018}). According to the SD scenario such an emission is expected in most SNe Ia (e.g. \citealt{Kasen2010}). However, such an emission is possible also in the DD scenario, as the ejecta collides with disk-originated matter (DOM; \citealt{Levanonetal2015, LevanonSoker2017, LevanonSoker2019}).  
\cite{Levanonetal2015} argued that in the frame of the DD scenario the presence of an early excess emission in only a small fraction of SNe Ia implies that in most cases the MED should be longer than tens of years to allow the DOM to disperse. 
    
(3) Another limit is on the ionisation radiation tens of thousands of years before the explosion of some SNe Ia, e.g., Tycho SN Ia. \cite{Woodsetal2017} find that the medium around the Tycho is not  ionised, and hence was not ionised before explosion. \cite{Woodsetal2018}, for Galactic SNRs, and \cite{Kuuttilaetal2019}, for Large Magellanic Cloud SNRs, constrain the pre-explosion ionisation of more SNe Ia. These studies put limits on the SD scenario. These results also put some limits on the ionisation from the merging WDs in the
DD scenario, as merging WDs might emit strong UV radiation 
(e.g., \citealt{TornambPiersanti2013}). 

Overall, I estimated \citep{Soker2018Rev} that the MED time of the DD scenario should be in  many cases $t_{\rm MED}({\rm DD}) \ga 10^5 \yr$, while in the SD scenario there are cases where $t_{\rm MED}({\rm SD}) \ga 10^7 \yr$. The DDet scenario with a WD companion and the WWC scenario allow for no MED time, and this is one of the problems of these scenarios \citep{Soker2018Rev}. In the CD scenario the MED time is built-in to the scenario, hence it is one of its advantages. 

In those scenarios where the binary system experiences a CEE, the time from the end of the CEE to explosion, $t_{\rm CEED}$, includes the MED time (section \ref{subsec:delay}). In the DD scenario the MED time is a small fraction of  $t_{\rm CEED}$, $t_{\rm MED}({\rm DD}) \ll t_{\rm CEED}$, while in  the CD scenario $t_{\rm MED}({\rm CD})=t_{\rm CEED}$, and its value might be up to billions of years if the CD scenario can allow for a long delay time (e.g., \citealt{Ilkov2012}). 

%=========================
\subsection{Common envelope to explosion delay (CEED) time}
\label{subsec:CEED time}
% ========================

The following considerations motivate me to define the CEED time and study the CEEDTD.

(1) The ejecta of the Kepler SNR Ia interact with a CSM (e.g., \citealt{Sankritetal2016}). The non detection of a giant star or a post-giant star (e.g., \citealt{Kerzendorfetal2014, Medanetal2017}) suggests that the CSM was blown during a CEE in the frame of either the CD scenario, the DD scenario, the CEE-wind channel of the SD scenario, or the DDet scenario. In the CEE-wind channel of the SD scenario for Kepler the remnant is a subdwarf B (sdB) star that is below observational limits \citep{MengLi2019}, while in the DDet scenario the remnant is a WD that might also be below observational limits and far from the center of Kepler SNR.   

(2) The mass of the CSM in the SNe Ia-CSM PTF11kx seems to be too large for the SD scenario with a giant donor \citep{Sokeretal2013}, and better fits mass ejection in a CEE. But I do note that \cite{MengPodsiadlowski2018} claim that their suggested CEE-wind channel of the SD scenario can account for a more massive CSM, such as that in PTF11kx. 

(3) In the DD scenario and in the DDet scenario with a WD donor the CEE forms the initial setting of two WDs. In the CD scenario the CEE forms the single WD merger product of the core and the WD companion. In the CEE-wind channel of the SD scenario the CEE ensures the right conditions to bring the WD to explode \citep{MengPodsiadlowski2017}. These suggest that an important time of evolution is the time from the end of the CEE to the explosion itself, i.e., $t_{\rm CEED}$. 
 
(4) The recent new derivations of parameters for the DTD \citep{FriedmannMaoz2018, Heringeretal2019} and the estimate of the fraction of SNe Ia-CSM \citep{Grahametal2019} allow an attempt to connect  the very short post-CEE time with times of $>1 \Gyr$.   

I attempt now such a derivation. 

%==========================================================
\section{Estimating the CEED time distribution (CEEDTD)}
\label{sec:Estimating CEED}
% ==========================================================

%================================================
\subsection{SN Ia rates from observations}
\label{subsec:Rate}
% ===============================================

 I turn to derive the CEEDTD in relative numbers, i.e., relative to the total number of SNe Ia. Unlike an expression for the CEEDTD in absolute values, i.e. per stellar mass, the relative CEEDTD is less sensitive to the uncertainties in the observationally derived absolute total number of SNe Ia per stellar mass and to new values of this absolute number that future observations might deduce. 
To crudely derive  the relative  CEEDTD I use the following expressions.
 
(1) \textit{Very long DTD.} I take equation (\ref{eq:dotN}) and substitute numbers from  \cite{Heringeretal2019} and \cite{FriedmannMaoz2018} (see discussion following equation \ref{eq:dotN}). This equation reads now 
\begin{equation}
\dot N_{\rm DTD}  = 0.19 N_{\rm Ia} F_1(t_{\rm i}) \left( \frac{t }{1 \Gyr} \right)^{-1.32} \Gyr^{-1},
\label{eq:dotN3}
\end{equation}
where $t_{\rm i}$ is the first time after star formation when SNe Ia occur, 
$F_1(t_{\rm i})=1.68 [ (t_{\rm i}/\Gyr)^{-0.32} - 13.7^{-0.32}]^{-1}$,
and with an uncertainty of $\alpha \simeq -1.32 \pm 0.2$. The scaling of $F_1(t_{\rm i})$ is such that $F_1(0.1 \Gyr)=1$. Integrating the rate in equation (\ref{eq:dotN3}) from $t_{\rm i}$ to $t=13.7 \Gyr$ gives a total SNe Ia number  of $N_{\rm Ia}$. For $t_{\rm i}=0.1 \Gyr$ for example, the maximum rate  (at $t=t_{\rm i}$) is $\dot N_{\rm DTD} = 4 N_{\rm Ia} \Gyr^{-1}$, while for $t_{\rm i}=0.04 \Gyr$ the maximum rate is $\dot N_{\rm DTD}= 9.5 N_{\rm Ia} \Gyr^{-1}$.
   
(2) \textit{SNe Ia inside planetary nebulae (SNIP).}
\cite{TsebrenkoSoker2015a} estimated that the fraction of SNe Ia that explode within a CSM, i.e., a planetary nebula or a remnant of a planetary nebula, is at least $\simeq 20 \pm 10 \%$ of all SNe Ia. These are termed SNIPs, including SNe Ia that explode inside proto-planetary nebulae \citep{Cikotaetal2017}. 
\cite{TsebrenkoSoker2015a} assumed that the dispersion time of the planetary nebulae is $t_{\rm SNIP} \approx 10^5 \yr$, but might be as long as $\approx 10^6 \yr$. 
I take here the dispersion time to be $t_{\rm SNIP} \approx 3 \times 10^5 \yr$. For example, for an expansion velocity of $10 \km \s^{-1}$ and an ejecta velocity of $10^4 \km \s^{-1}$ the ejecta will interact with the CSM at a SN age of  $\simeq 300 \yr$. 
 
As an indication for the presence of a CSM \cite{TsebrenkoSoker2015a} took the presence of two opposite protrusions termed `Ears' in the SNR (see also \citealt{Chiotellisetal2016}). 
They find that out of their 13 SNRs Ia two posses ears and 4 maybe possess ears. From this they estimated that $\simeq 15-45 \%$ of the SNRs Ia are SNIPs. 
However, the SNR Ia N103B that they did not list as a SNIP does interact with a CSM 
 (e.g., \citealt{Williamsetal2018}). 
If I take the two SNRs that are known to interact with a CSM from the list of 13 SNRs, Kepler and N103B, I find the fraction of SNIPs to be $\approx 15 \%$. 
  
 I note that in principle there are two other possibilities for the formation of SNRs with two opposite protrusions. The first one is the formation of a CSM by a giant companion to the WD in the frame of the SD scenario, without going through the planetary nebula phase. This cannot work for the Kepler SNR as there is no giant companion there (e.g., \citealt{Kerzendorfetal2014, Medanetal2017}). The second possibility is that there is an ISM close to the SN, and the interaction of an axisymmetrical explosion with two opposite clumps (jets) form the two protrusions \citep{TsebrenkoSoker2013}. It is not clear if a regular ISM can form such a massive CSM. For the above reservations, I consider in the present study only the CEE channel to form the dense CSM. 
  
Overall, I take for the SNIP fraction out of all SNe Ia and for the planetary nebula dispersion time $f_{\rm SNIP} \simeq 15-20 \%$ and $t_{\rm SNIP} \simeq 3 \times 10^5 \yr$, respectively, from which I estimate the average SN Ia rate in the time interval $0< t_{\rm CEED} < 3 \times 10^5 \yr$ to be 
\begin{equation}
%\overline
{\overline {\dot N}}_{\rm SNIP} 
= \frac{f_{\rm SNIP} N_{\rm Ia}}{t_{\rm SNIP}} \approx (100-1000) N_{\rm Ia} \Gyr^{-1}.
\label{eq:SNIPrate}
\end{equation}
This rate is much larger than the rate that equation (\ref{eq:dotN3}) gives for $0< t_{\rm CEED} < 3 \times 10^5 \yr$. 
For example, if the time from star formation to CEE is $0.1 \Gyr$ then the time in equation (\ref{eq:dotN3}) is $t=0.1 \Gyr + t_{\rm CEED}$.
I conclude therefore, that equation (\ref{eq:dotN3}) cannot be used as is to give the SN Ia rate short times after the CEE. 

(3) \textit{SNe Ia-CSM.}
There are SNe Ia that show signatures of interaction with CSM within months after explosion, e.g.,  PTF11kx \citep{Dildayetal2012} and SN~2015cp \citep{Grahametal2019}. Such SNe Ia-CSM are very rare (e.g., \citealt{Szalaietal2019}). From their detection of CSM interaction 686 days after explosion \cite{Grahametal2019} determine the maximum inner
radius of the CSM to be $R_{\rm CSM} \la 10^{17} \cm$. 
For a CSM expansion velocity of $10 \km \s^{-1}$ the time from the end of the CEE (assuming the CSM was formed in a CEE) to explosion is $t_{\rm CSM} \la 3000 \yr$.  
\cite{Grahametal2019} further estimate that the fraction of SNe Ia-CSM is $f_{\rm CSM} < 0.06$ of all SNe Ia. 
 
I crudely estimate the SNe explosion rate within the CSM interaction time by taking  $t_{\rm CSM} \approx 1000 - 3000 \yr$ and $f_{\rm CSM} \approx 0.03-0.05$. This gives for the average SN Ia rate at $t_{\rm CEED}=t_{\rm CSM}  \approx 1000 - 3000 \yr$ 
\begin{equation}
%\overline
{\overline {\dot N}}_{\rm CSM} 
= \frac{f_{\rm CSM} N_{\rm Ia}}{t_{\rm CSM}} \approx (10^4-5 \times 10^4 )  N_{\rm Ia} \Gyr^{-1}.
\label{eq:CSMrate}
\end{equation}
I note that the SN Ia fraction $f_{\rm SNIP} \simeq 0.15-0.2$ includes the fraction $f_{\rm CSM}\la 0.06$. Namely the fraction of SNe Ia that explode inside extended planetary nebulae but show no interaction within few years from explosion is $f_{\rm SNIP}-f_{\rm CSM} \approx 0.1-0.2$. 

%================================================
\subsection{A crude plausible short CEED time distribution}
\label{subsubsec:CEED}
% ===============================================

Equations (\ref{eq:dotN3}), (\ref{eq:SNIPrate}), and (\ref{eq:CSMrate}) show that the SN Ia rates at short times after the  CEE, i.e., the SNe Ia-CSM and the SNIPs, require a different expression for their rate, and that the time from the CEE to explosion, $t_{\rm CEED}$, is a better measure than the time from star formation. 
The two rates of the two populations, of SNIPs and of SNe Ia-CSM, do not allow to derive an expression. 
I make two more assumptions to derive a plausible expression, but it is definitely not a unique expression. It only serves to emphasise some properties of these populations. 

(1) I assume that the time from the end of the CEE to explosion, $t_{\rm CEED}$ of a specific system is sensitive to a parameter $\aleph$ (pronounced `aleph') according to 
\begin{equation}
t_{\rm CEED} \propto \aleph ^{\eta} \quad \rightarrow \quad  \frac{d \aleph}{d t_{\rm CEED}}
\propto  \left( t_{\rm CEED} \right)^{\eta^{-1}-1} ; \quad \eta \gg 1, 
\label{eq:tauexp}
\end{equation}
and that $\aleph$ decreases with time. Let the formation of systems to be exploded, like WD binary systems in the DD scenario or single WDs in the CD scenario, be distributed in a weakly-dependent manner on $\aleph$ at the end of the CEE. Namely,
\begin{equation}
\left( \frac {dN_{\rm Ia}}{d \aleph} \right)_{t_{\rm CEED}=0}  \propto \aleph ^{\epsilon}; \quad   -1 \la \epsilon \la 1.
\label{eq:npe}
\end{equation}
From equations (\ref{eq:tauexp}) and (\ref{eq:npe}) one gets
\begin{equation}
\begin{aligned}
\dot N_{\rm Ia}  =   &
\frac{d N_{\rm e}}{d \aleph} \frac{d \aleph}{d t_{\rm CEED}}
\propto \left( t_{\rm CEED} \right)^{\epsilon/\eta} 
\left( t_{\rm CEED} \right)^{\eta^{-1}-1} \\ &
\simeq  \left( t_{\rm CEED} \right)^{-1}, \quad {\rm for} \quad \eta \gg 1.
\label{eq:Neexp}
\end{aligned}
\end{equation}

This derivation is not new, e.g., \cite{Greggio2005}. For the DD scenario, for example, the orbital decay is due to gravitational radiation, so the parameter is the orbital separation, i.e.,  $\aleph \rightarrow a$ with $\eta=4$ and $\epsilon \simeq -1$ and one obtains $d N_{\rm Ia}/d t_{\rm CEED} \propto (t_{\rm CEED})^{-1}$ (e.g., \citealt{Maoz2010}). But for the specific populations I focus on the parameter might be another one, e.g, the angular momentum of the WD that was formed by the WD-core merger in the CD scenario. 

(2) The second assumption I make is that the rate of equation (\ref{eq:Neexp}) is applicable in a relatively short time range of $t_1 \la t_{\rm CEED} \la t_2$, where $t_1 \approx 1000 \yr$ and $t_2 \approx 10^6 -10^7 \yr$. The upper limit is similar to what \cite{MengPodsiadlowski2018} argue for in the CEE-wind channel of the SD scenario. 
 
I can use now the two rates given in equations (\ref{eq:SNIPrate}) and (\ref{eq:CSMrate}) with the above time limit, to write for the rate shortly after the CEE,  i.e., for the CEEDTD 
\begin{equation}
\begin{aligned}
& \dot N_{\rm Ia,short}  \approx 10^{3.7 \pm 0.2} N_{\rm Ia}
\left( \frac{t_{\rm CEED}}{10^4 \yr} \right)^{-1} \Gyr^{-1} ,
\\ &  {\rm for} \quad  10^3 \yr \approx t_1 \le t_{\rm CEED} \le t_2 \approx 3 \times 10^6 \yr, 
\label{eq:NeexpFinal}
\end{aligned}
\end{equation}
with large uncertainties in the time range and in the rate itself.  
  
Despite the large uncertainties in expression (\ref{eq:NeexpFinal}), both in its form and in its numerical values, it emphasises two properties of the SN Ia population that takes place shortly, within $\approx 10^3-10^6 \yr$, after the CEE, i.e., SNIPs and SNe Ia-CSM.

(1) Integrating equation (\ref{eq:NeexpFinal}) over the time span and for the lower value coefficient $10^{3.7-0.2}$  gives a total SNe Ia population of $N_{\rm Ia,short} \approx 0.03 N_{\rm Ia} \ln (t_2/t_1)$. 
For $t_2=3000t_1$ this gives $N_{\rm Ia,short} \approx 0.25 N_{\rm Ia}$ and for $t_2=300t_1$ this gives $N_{\rm Ia,short} \approx 0.18 N_{\rm Ia}$. For the upper value of $10^{3.7+0.2}$ the values are $2.5$ larger. 
Over all I find $N_{\rm Ia,short} \approx {\rm few} \times 0.1 N_{\rm Ia}$. 

(2) If we would have continue equation (\ref{eq:dotN}) to short times down to $t=1000 \yr$, it would be always larger than the rate given by equation (\ref{eq:NeexpFinal}) for $t \la 1 \Gyr$.  This hints that the physical processes that determine the delay time to explosion shortly after the CEE are not identical to those that determine the delay time at very long times.
Due to the very large uncertainties this conclusion is only a tentative one. 
 
%================================================
\subsection{Observational tests}
\label{subsubsec:Tests}
% ===============================================

  The tentative conclusion that I derived above requires more observational test, as I suggest below. The first one is to follow SNe Ia for a long time, years to tens of years, after explosion. In cases where the CSM is at distances of $\la 1 \pc$ the ejecta-CSM collision within several years might cause re-brightening in different bands, e.g., optical and UV (e.g., \citealt{Grahametal2019}), X-ray, and radio. The analysis of this study suggests that $\approx 10\%$ of all SNe Ia should experience ejecta-CSM interaction within 30 years. We might detect signatures of interaction in nearby SNe Ia. The analysis of re-brightening tens of years after explosion deserves a study of its own. 

 Another observational test might be light-echo. The CSM might contain large mass of dust that might reflect a large portion of the SN light. \cite{Maund2019}, for example, suggests that the re-brightening of the type IIb SN~2011dh in M51 that took place few years after explosion might come from light echo. 
Some SNe Ia also show light echo (e.g, \citealt{Graur2019}). When resolved, I predict that in some cases the morphology of the echoing dust will have axisymmetrical structure, such as that of planetary nebulae. In cases where the CSM is large, interaction with the ISM might change completely its geometry. 

 Finally, I predict that in the local group, where we can easily detect planetary nebulae, in $\approx 10 \%$ of SNe Ia we might detect a pre-explosion planetary nebula. The planetary nebula might be very faint, so the detection is not easy. Although the chance of detecting pre-explosion planetary nebulae is very low (as there are not many SNe Ia in the local group), it is not zero.

%==========================================================
\section{SUMMARY}
\label{sec:summary}
% ==========================================================

The goal of the present study is a derivation of a crude SNe Ia rate,  the CEEDTD,  as function of the time $t_{\rm CEED}$ after the CEE that, according to my assumption, forms the progenitors of most SNe Ia.
For that, the present study is relevant to the CD scenario, the DD scenario, the DDet scenario with a WD companion, and to the CEE-wind channel of the SD scenario.   
 
While the usual DTD refers to a long time after star formation (equations \ref{eq:dotN} and \ref{eq:dotN3}), in this study I focused on the rate of SNe Ia that interact with a CSM within months after explosion, so called SNe Ia-CSM (equation \ref{eq:CSMrate}), and the rate of SNe Ia that interact with a CSM that might have been a planetary nebula, so called SNIPs (equation \ref{eq:SNIPrate}). 
To derive a plausible expression for the CEEDTD (equation \ref{eq:NeexpFinal}) I made two assumptions (section \ref{subsubsec:CEED}). This expression is crude (and not unique).

Despite the very large uncertainties in the parameters and time span of equation (\ref{eq:NeexpFinal}) it emphasises the conclusions of this study. 
\begin{enumerate}
  \item There is a large population of SNe Ia, $\approx {\rm few} \times 0.1$ of all SNe Ia, that explode a short time, within $t_{\rm CEED} \approx 10^6 \yr$ (and possibly up to $t_{\rm CEED} \approx 3 \times 10^6 \yr$), after the CEE.
\item The expression for the SNe Ia rate as a function of time after the CEE, the CEEDTD, cannot be the one that is used for DTD long after star formation. 
\item The previous conclusion hints that the physical processes that determine the short delay time from the CEE to explosion, i.e., of the SNe Ia-CSM and of SNIPs that occur at $t_{\rm CEED} \la 10^6 \yr$, are different (at least to some extend) from those that determine the DTD at long time scales of $t_{\rm CEED} \ga 10^7 \yr$. This very tentative conclusion deserves deeper studies. 
\end{enumerate}

 I thank an anonymous referee for useful comments.  
This research was supported by the Israel Science Foundation.

\end{document}